# Single crystals of superconducting $SmFeAsO_{1-x}F_y$ grown at high pressure

N D Zhigadlo[a], S Katrych[a], Z Bukowski[a], S Weyeneth [b], R Puzniak[c], J Karpinski[a]

[a]*Laboratory for Solid State Physics, ETH, 8093 Zürich, Switzerland*
[b]*Physik-Institut der Universität Zürich, Winterthurerstrasse 190, 8057 Zürich, Switzerland*
[c]*Institute of Physics, Polish Academy of Sciences, Aleja Lotnikow 32/46, 02-668 Warsaw, Poland*

E-mail: zhigadlo@phys.ethz.ch  and karpinski@phys.ethz.ch

**Abstract.** Single crystals of $SmFeAsO_{1-x}F_y$ of a size up to 120x100 μm$^2$ have been grown from NaCl/KCl flux at pressure of 30 kbar and temperature of 1350-1450 °C using cubic anvil high-pressure technique. Superconducting transition temperature of the obtained single crystals varies between 45 and 53 K. Obtained crystals are characterized by full diamagnetic response in low magnetic field and by high critical current density in high magnetic field. Structure refinement has been performed on single crystal. Differential thermal analysis investigations at 1 bar Ar pressure show decomposition of $SmFeAsO_{1-x}F_y$ at 1302 °C.

## 1. Introduction

The recent discovery of superconductivity in a family of quaternary oxypnictides with the general formula LnFeAsO (where Ln = La, Ce, Pr, Nd, Sm, Gd) has caused excitement in the scientific community [1-8]. The crystal structure of PrFeAsO was reported by Quebe *et al.* [9]. These compounds crystallize with the tetragonal layered ZrCuSiAs structure in the space group of P4/nmm, which has a structure of alternating LnO and FeAs layers that are electrically charged and can be represented as $(LnO)^{+\delta}(FeAs)^{-\delta}$ (Fig. 1). Covalent bonding is dominant in the layers, while between layers dominates ionic bonding. The charge carriers in these compounds are electrons confined to the FeAs layers. The LnO layer serves as a "charge reservoir", when doped with electrons. This can be done by partial substitution of $O^{2-}$ by fluorine $F^-$ [2-5] or by generation of oxygen vacancies at high pressure of 60 kbar [6]. The highest critical temperatures above 50 K have been observed for $LnFeAsO_{1-x}F_x$ with Ln = Sm, Gd, Pr and Nd. Both methods of doping lead to a decrease in lattice parameters, which indicates that high pressure can promote such reactions. The presence of two structural blocks, namely conducting FeAs layers and "charge reservoir" LnO layers, recalls the high-$T_c$ cuprates. One can expect, similar to cuprates, strong anisotropy of superconducting properties such as upper critical fields, coherence length, and penetration depth. In fact investigations of penetration depth anisotropy with the torque technique show temperature dependent anisotropy varying from γ = 8 at $T \leq T_c$ to γ = 23 at $T = 0.4\ T_c$ [7]. Penetration depth investigations using a radio frequency tunnel diode oscillator technique performed on single crystals show an exponential temperature dependence, indicating a fully gapped pairing state [8]. The knowledge of the anisotropic properties is crucial for the understanding of the mechanism of superconductivity in this family of compounds and for their potential applications.

The two techniques have been used for the synthesis of polycrystalline samples: the low-pressure quartz ampoule method [1-3] and high-pressure synthesis [4-6]. As a precursor a mixture of LnAs, FeAs, $Fe_2O_3$, Fe, and $LnF_3$ is usually used. In the low-pressure method, the necessary reaction temperature range of 1150-1250 °C is at the limit of application of quartz ampoules due to reaction with precursor, especially with fluorine vapours. In the high-pressure method the precursor mixture is placed in a BN crucible and synthesized at a pressure of 30 - 60 kbar at temperatures of 1250-1350 °C for several hours. Sintered polycrystalline samples with a micrometer grain size have been obtained. Until now most of the physical measurements have been performed on polycrystalline samples obtained in one of these ways. As the mechanism of superconductivity in these pnictide oxides is unknown, single crystals are necessary for the investigations of intrinsic anisotropic properties such as upper critical fields, coherence length or penetration depth. Studies on single crystals are crucial for spectroscopic techniques such as scanning tunnelling spectroscopy, angle-resolved photoemission spectroscopy, point contact spectroscopy and optical investigations. In the low-pressure method for the iron arsenide synthesis NaCl/KCl flux has been reported as a mineralizer, which enhances the formation of the quaternary compounds [9]. The authors emphasized that at the conditions of synthesis (800 °C) only minor amounts of the metallic component were dissolved. We decided to apply this method at high-pressure, which allowed us to use a higher temperature. Up to submission of our paper, the growth of superconducting free single crystals of $LnFeAsO_{1-x}F_x$ has not been reported. The authors of [10] reported synthesis and magnetic properties of $Nd(O_{0.9}F_{0.1})FeAs$ sintered sample containing large crystallites obtained by a similar high-pressure method. In this paper we report on the growth and properties of $SmFeAsO_{1-x}F_y$ single crystals.

## 2. Experimental details

For the synthesis of polycrystalline samples and single crystals of $SmFeAsO_{1-x}F_y$ we used cubic anvil high-pressure technique which has been used in our laboratory for the growth of $MgB_2$ single crystals and other superconductors. Polycrystalline samples of nominal composition $SmFeAsO_{0.8}F_{0.2}$ were prepared using SmAs, FeAs, $Fe_2O_3$, Fe and $SmF_3$ powders as starting materials. For the growth of single crystals we used the same components and NaCl/KCl flux. The precursor to flux ratio varies between 1:1 and 1:3. The mixing and grinding of precursor powders and pressing pellets have been performed in a glove box due to toxicity of arsenic. Pellets containing precursor and flux were placed in BN crucible inside a pyrophyllite cube with a graphite heater. The six tungsten carbide anvils generated pressure on the whole assembly. In a typical run, a pressure of 3 GPa was applied at room temperature. While keeping pressure constant, the temperature was ramped up within 1 h to the maximum value of 1350-1450 °C, maintained for 4-10 h and decreased in 5-24 h to room temperature for the crystal growth. For the synthesis of polycrystalline samples the maximum temperature was maintained for 2-4 h, followed by quenching. Then pressure was released, sample removed and in the case of single crystal growth NaCl/KCl flux dissolved in water. One has to mention that such high-pressure experiments have to be performed very carefully, because an explosion during heating due to increased pressure in the sample container could lead to contamination of the whole apparatus with arsenide compounds. Differential thermal analysis (DTA) was carried out in a Perkin Elmer DTA 7 analyzer using $Al_2O_3$ crucibles in flowing Ar with a heating rate of 5 °C/min up to 1600 °C. The magnetization was measured with a Quantum Design SQUID magnetometer. Structural investigations were done using a diffractometer equipped with a charge-coupled device (CCD) detector (Xcalibur PX, Oxford Diffraction). Data reduction and analytical absorption correction were performed using the program CrysAlis [11]. The crystal structure was determined by a direct method and refined on $F^2$, employing the programs SHELXS-97 and SHELXL-97 [12, 13].

## 3. Results

In order to determine temperature limits for the crystal growth in ambient pressure, DTA investigations at 1 bar Ar pressure were performed. For this experiment a polycrystalline sample of $SmFeAsO_{1-x}F_y$ obtained at high pressure was used. The results of the run are shown in Fig. 2. One can notice two endothermic peaks corresponding to two reactions. The first one, with an onset at 993 °C and a maximum at 1020 °C, corresponds to melting of FeAs, which was an impurity in our $SmFeAsO_{1-x}F_y$ sample. The second one, with an onset at 1302 °C, corresponds to incongruent melting of $SmFeAsO_{1-x}F_y$. This practically excludes increasing temperature for the crystal growth at ambient pressure much above 1250 °C used usually for the synthesis of polycrystalline samples. Samples obtained at this temperature have very fine micrometer size grains. High-pressure extends the stability range of the compound to higher temperatures, which allows application of higher temperature for the crystal growth. This increases both the size of the grains and the solubility in the flux. As a result of the crystal growth experiments $SmFeAsO_{1-x}F_y$ plate-like crystals were obtained. Crystals with a size about 120x100x20 $\mu m^3$ grown in this way are presented in Fig. 3. One can see remnants of the solidified flux on the surface of some crystals.

With the aim of growing single crystals suitable for physical measurements, we carried out a systematic investigation of the parameters controlling the growth of crystals, including temperature, pressure, composition, reaction time and heating/cooling rate.

One of the problems of crystal growth under high-temperature and high-pressure conditions is that the density of sites for nucleation is high and it is difficult to control nucleation to produce a small number larger single crystals and not many small crystals. This is also reflected on the quality of grown crystals, therefore many of them have irregular shapes and they form clusters of several crystals. The solubility of SmFeAsO$_{1-x}$F$_y$ in NaCl/KCl flux is very low, which results in a small crystal size. Unfortunately, so far no alternative solvent for this compound has been found. Several parameters which influence the quality of crystals can be specified: (i) starting composition of precursor; (ii) the precursor to flux ratio; (iii) time of dwelling and cooling, etc. For example, increase of the dwelling time at 1380 °C up to 10 h and increase in the cooling time from this temperature down to 1000 °C up to 50 h results in single crystals with a size of 120x120 μm$^2$. The existence of parasitic phases such as FeAs also has a significant affect on the growth mechanism and appropriate doping. A high precursor to flux ratio prevents growth of larger crystals because of a lack of space for the growth of individual grains. Further experiments with different kind of fluxes are planned to optimize growth conditions and to grow larger and optimally doped crystals.

*3.1. Crystal structure*

All atomic positions were found by a direct method. The structure was refined without any restrains. Oxygen and fluorine atoms which occupy the same site are impossible to distinguish by x-ray diffraction so they were treated during refinement as one atom. The results of the structure refinement are presented in Tables 1 and 2.

The results are in good agreement with the published data for the PrFeAsO [9]. According to the reflection conditions for the space group P4/nmm ($hk0 = 2n$) the systematic absences occur only for the $hk0$ reciprocal section (Fig. 4). Structural analysis revealed overall occupancy in the O(F) site to be considerably lower than 100 %, equal to 0.86. Therefore, it is possible that except for F doping there is also an additional electron doping due to O vacancies. However, the accuracy of the determination of the oxygen or fluorine occupancy factor is very low because of the presence of such heavy atoms like As, Fe and Sm in the unit cell. The residuals $R_1$ and w$R_2$ as well as goodness of fit $S$ show small difference for the oxygen occupancy of 86 % and 100 % (Tab. 3). The minimal residuals correspond to the O(F)$_{occ.}$ of 86 ± 3 at %. Neutron diffraction data could be helpful for clarifying this point.

**Table 1.** Crystal data and structure refinement for the SmFeAsO$_{0.86-x}$F$_x$

| | |
|---|---|
| Empirical formula | SmFeAsO$_{0.86-x}$F$_x$ |
| Temperature, K | 295(2) |
| Wavelength, Å | 0.71073/MoK$_\alpha$ |
| Crystal system, space group, Z | Tetragonal, $P4/nmm$, 2 |
| Unit cell dimensions, Å | $a$= 3.93390(10), $c$= 8.4684(6), |
| Volume, Å$^3$ | 131.053(10) |
| Calculated density, g/cm$^3$ | 7.49 |
| Absorption correction type | analytical |
| Absorption coefficient, mm$^{-1}$ | 39.914 |
| $F$(000) | 257 |
| Crystal size, μm$^3$ | 70 x 30 x 10 |
| Theta range for data collection | 4.81 to 37.19 deg |
| Index ranges | -5=<h<=6, -6<=k<=6, -14<=l<=13 |
| Reflections collected/unique | 867/235 R$_{int.}$= 0.0399 |
| Completeness to 2theta | 97.5 % |
| Refinement method | Full-matrix least-squares on $F^2$ |
| Data/restraints/parameters | 235/0/12 |
| Goodness-of-fit on $F^2$ | 1.017 |
| Final R indices [I>2sigma(I)] | $R_1$ = 0.0323, w$R_2$ = 0.0708 |
| $R$ indices (all data) | $R_1$ = 0.0461, w$R_2$ = 0.0746 |
| Δρ$_{max}$ and Δρ$_{min}$,(e/Å$^3$) | 3.594 and -2.454 |

**Table 2.** Atomic coordinates and equivalent isotropic and anisotropic displacement parameters [Å$^2$ x 10$^3$] for the SmFeAsO$_{0.86-x}$F$_x$

| Atom | Site | $x$ | $y$ | $z$ | $U_{iso}$ | $U_{11}$=$U_{22}$ | $U_{33}$ |
|---|---|---|---|---|---|---|---|
| Sm | 2c | -1/4 | -1/4 | 0.1411(1) | 11(1) | 10(1) | 12(1) |
| Fe | 2b | 1/4 | 3/4 | 1/2 | 10(1) | 10(1) | 10(1) |
| As | 2c | 1/4 | 1/4 | 0.3391(2) | 10(1) | 9(1) | 12(1) |
| O(F) | 2a | 1/4 | 3/4 | 0 | 11(3) | 10(3) | 15(5) |

$U_{iso}$ is defined as one third of the trace of the orthogonalized $U_{ij}$ tensor. The anisotropic displacement factor exponent takes the form: -2π$^2$ [ ($h^2a^2U_{11}$ + ... + 2$hka*b*U_{12}$]. For symmetry reasons $U_{23}$=$U_{13}$=$U_{12}$=0.

**Table 3.** Residuals and goodness of fit for the different O(F) occupations in the SmFeAsO$_{1-x}$F$_y$

| O(F)$_{occ.}$, at. % | $R_1$ | w$R_2$ | $S$ |
|---|---|---|---|
| 86 | 0.0461 | 0.0746 | 1.017 |
| 100 | 0.0474 | 0.0771 | 1.050 |
| 0 | 0.0849 | 0.1554 | 2.120 |

*3.2. Superconducting properties*

Magnetic measurements of SmFeAsO$_{1-x}$F$_y$ crystals show that $T_c$ varies between 45 and 53 K dependently on doping. Figure 5a shows the temperature dependence of magnetic susceptibility measured on a collection of single crystals from one batch. The measurements were carried out in a magnetic field of 5 Oe on heating after zero-field cooling and then on cooling in a field. The superconducting volume fraction is large enough to constitute bulk superconductivity. The maximum $T_c$ reported for this compound is 55 K. A lower $T_c$ indicates non-optimal doping. The relatively broad width of transition is caused by the difference in $T_c$ between crystals from the same batch. Temperature and compositional gradients in the crucible may lead to difference in the F content of crystals and differences in $T_c$.

The temperature dependence of magnetic moment measured in a magnetic field parallel to the *c*-axis for one single crystal from the same batch, with dimensions of about 0.09-0.12 x 0.09 x 0.006 mm$^3$ and a mass of about 0.41 μg, is presented in Figure 5 b. A sharp transition to the superconducting state is characteristic of single crystals. A transition temperature of 48 K indicates that the crystal is underdoped. The value of zero-field cooled magnetic moment reflects the full diamagnetic response of the crystal studied. A small ratio of field cooled to zero-field cooled magnetization is characteristic of superconductors with strong pinning. The hysteresis loop measured at 5 K in magnetic field up to 7 T parallel to the crystal *c*-axis is presented in Figure 6 a. The wide loop, with a width almost independent on the field, proves the high critical current density of the crystal. The critical current density estimated from the width of the hysteresis loop is of the order of 10$^6$ A/cm$^2$ in the full field range investigated (see, Figure 6 b). The slight increase in critical current density for higher magnetic field may indicate an increase in the effectiveness of pinning centres with increasing magnetic field.

**4. Conclusions**

Single crystals of SmFeAsO$_{1-x}$F$_y$ superconductor have been grown using the high-pressure cubic anvil technique. The crystals have a plate-like shape with size up to 120 μm and are superconducting below 45-53 K. The crystal structure of SmFeAsO$_{1-x}$F$_y$ refined from single crystal x-ray diffraction data shows incomplete occupancy of the O(F) position. Magnetic measurements in a field up to 7 T show a relatively high critical current density of 10$^6$ A/cm$^2$ almost independent of the field.

**Acknowledgments**

This work was supported by the Swiss National Science Foundation through the NCCR pool MaNEP.

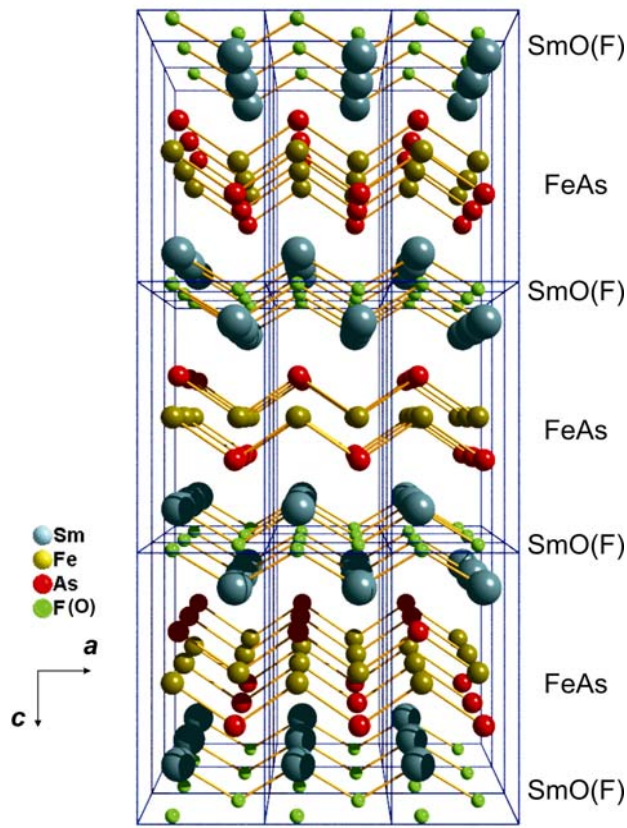

**Figure 1.** Schematic representation of the 3x3x3 unit cells of SmAsFeO$_{1-x}$F$_y$ along the *a* direction.

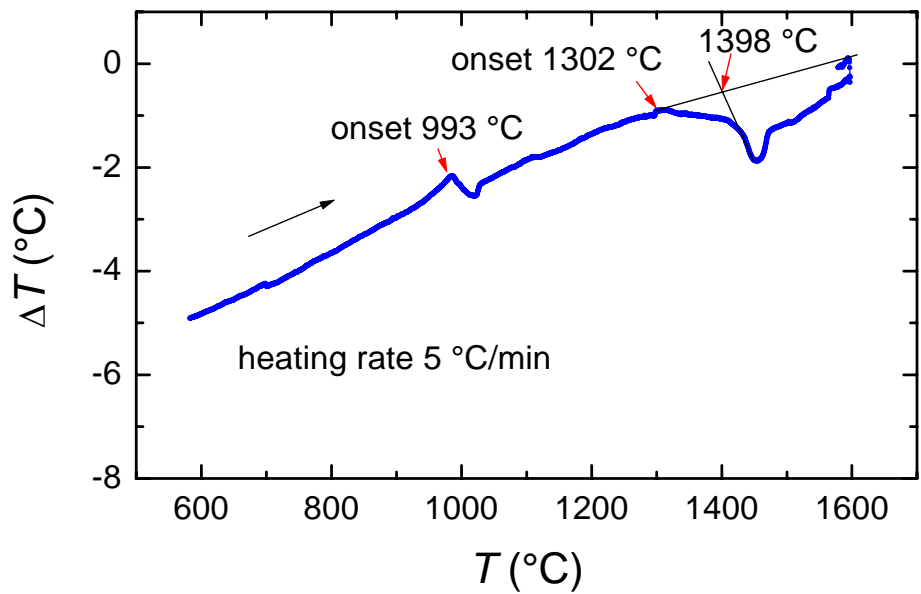

**Figure 2.** Differential thermal analysis performed in 1 bar Ar on a SmAsFeO$_{1-x}$F$_y$ polycrystalline sample showing decomposition the onset of decomposition (incongruent melting) at 1302 °C.

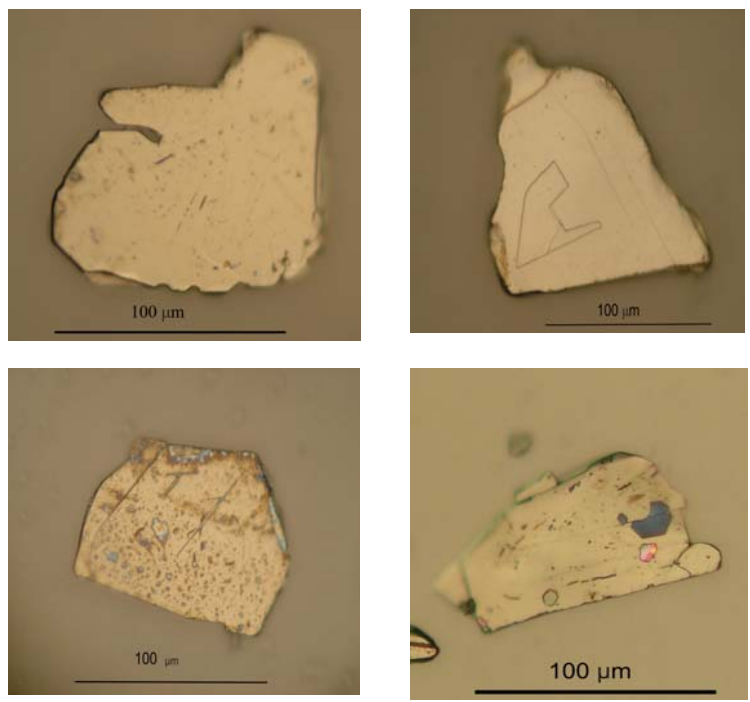

**Figure 3.** SmFeAsO$_{1-x}$F$_y$ single crystals grown from NaCl/KCl flux at high pressure. The scale is shown at the bottom.

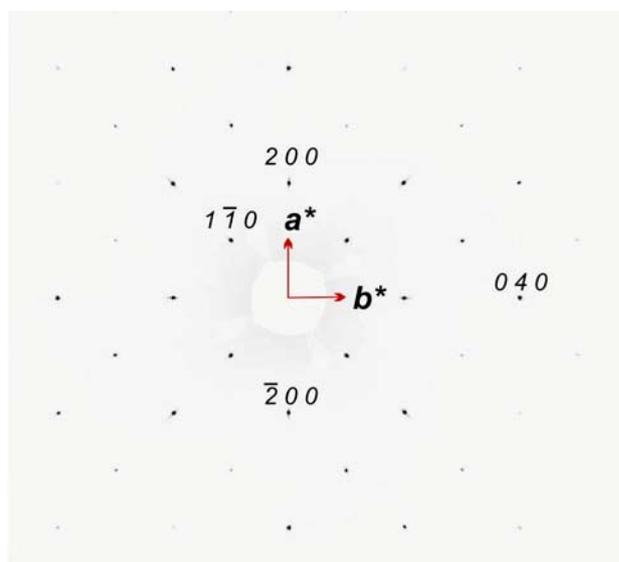

**Figure 4.** The reconstructed *hk0* reciprocal space section of the SmFeAsO$_{1-x}$F$_y$ single crystal.

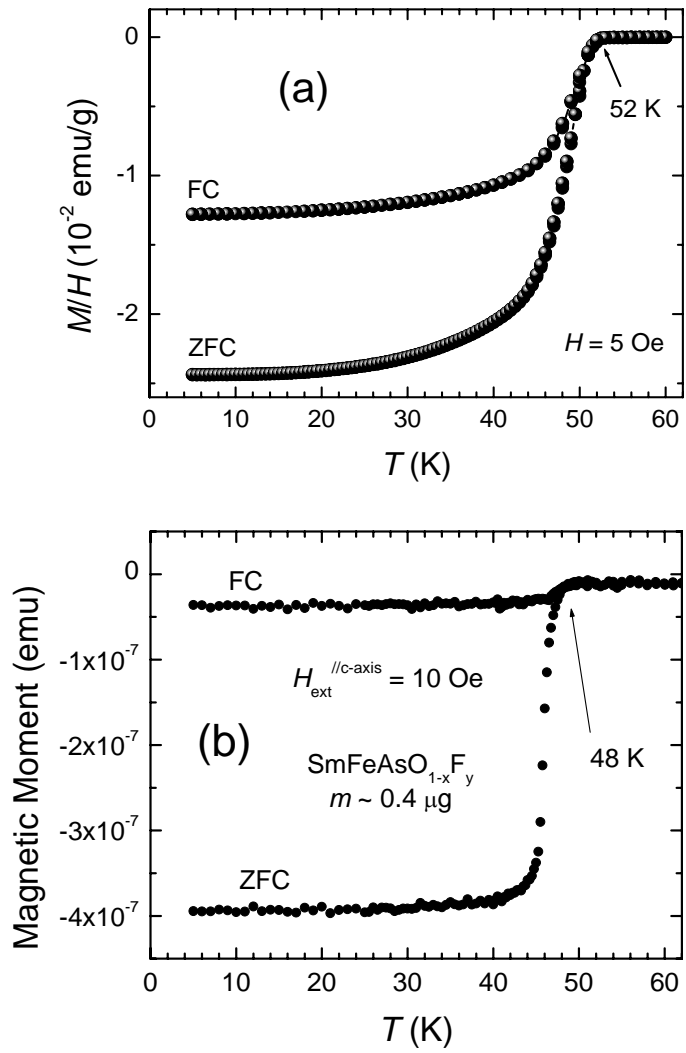

**Figure 5 (a)**: Temperature dependence of magnetic susceptibility measured on a collection of randomly oriented SmFeAsO$_{1-x}$F$_y$ crystals in an applied field of 5 Oe.
**(b)**: Temperature dependence of the magnetic moment measured on one single crystal in an applied field of 10 Oe. ZFC and FC mean zero-field cooling and field cooling curves, respectively.

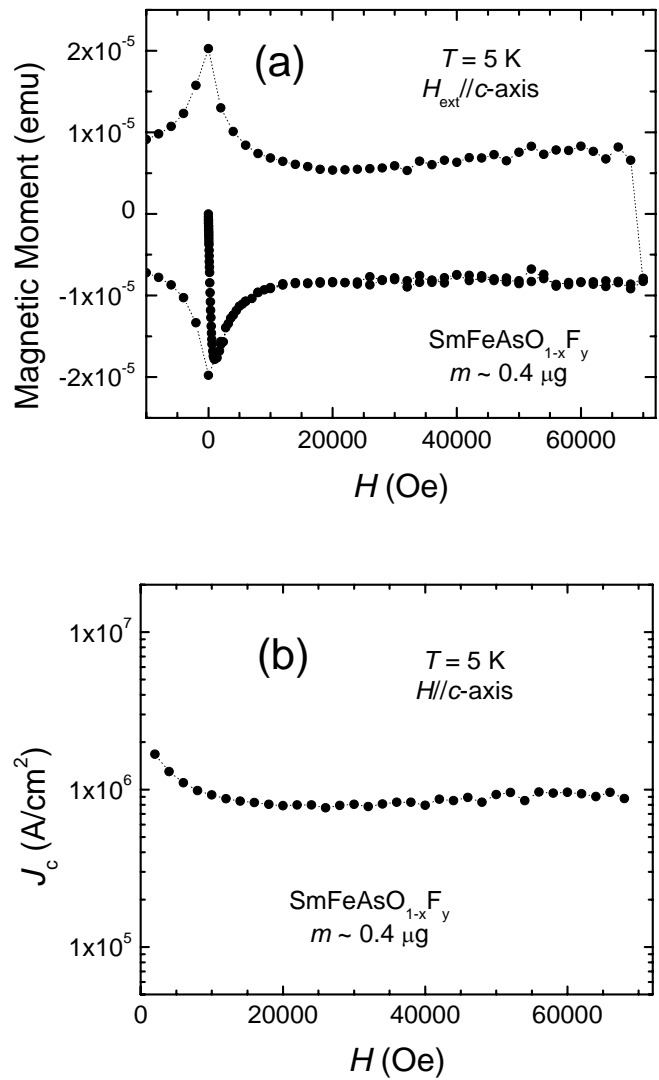

**Figure 6 (a)**: Magnetic hysteresis loop measured on a single crystal at 5 K in a field up to 7 T parallel to the *c*-axis. **(b)**: Critical current density calculated from the width of the hysteresis loop.